\documentclass[conference]{IEEEtran}
\usepackage[T1]{fontenc}
\IEEEoverridecommandlockouts
\usepackage{cite}
\usepackage{amsmath,amssymb,amsfonts}
\usepackage{algorithmic}
\usepackage{graphicx}
\usepackage{textcomp}
\usepackage{afterpage}
\usepackage[hyphens]{url}
\usepackage{hyperref}
\usepackage{xcolor}
\usepackage{hyperref}
\usepackage{amsmath}    
\usepackage{amssymb}    
\usepackage{amsthm}     
\usepackage{bm}         
\usepackage{booktabs}

\usepackage{tikz}
\usepackage{pgfplots}
\usepackage{caption}
\usetikzlibrary{shapes.geometric, arrows, positioning}

\usepackage{titlesec}

\titlespacing*{\section}{0pt}{0.5ex}{0.3ex}
\titlespacing*{\subsection}{0pt}{0.4ex}{0.2ex}
\titlespacing*{\subsubsection}{0pt}{0.3ex}{0.2ex}

\usepackage{hyperref}   
\hypersetup{
    colorlinks=true,
    linkcolor=blue,
    citecolor=blue,
    urlcolor=blue
}

\usepackage{tabularx}
\usepackage{caption}   
\usepackage{subcaption} 
\usepackage{multirow}   
\usepackage{siunitx}    
\def\BibTeX{{\rm B\kern-.05em{\sc i\kern-.025em b}\kern-.08em
    T\kern-.1667em\lower.7ex\hbox{E}\kern-.125emX}}
\begin{document}

\title{Machine Learning-Based Delivery Time Prediction for Low-Carbon Food Delivery}
\vspace{-2mm}
\author{
\IEEEauthorblockN{Ahmad M. Almasabi\IEEEauthorrefmark{1}, Mariem Belhor\IEEEauthorrefmark{2} and Omar Alam\IEEEauthorrefmark{3}}
\IEEEauthorblockA{
\IEEEauthorrefmark{1}College of computer Science \& Information Systems, Najran University, Najran city, Saudi Arabia\\
\IEEEauthorrefmark{2} University of Picardie Jules Verne, Laboratoire des Technologies Innovantes (LTI) UR 3899, Amiens, France\\
\IEEEauthorrefmark{3}Trent University, Peterborough, Ontario, Canada\\
Emails:  amalkheder@nu.edu.sa, mariem.belhor@u-picardie.fr, omaralam@trentu.ca}
}
\maketitle

\begin{abstract}
The increasing demand for online food delivery services has significantly impacted urban logistics, raising critical challenges related to operational efficiency and environmental sustainability. In this context, accurate delivery time prediction plays a key role in improving service quality while supporting the transition toward low-carbon delivery systems. This paper first provides a comprehensive analysis of sustainable and low-carbon delivery modes for urban food distribution. The study then develops and evaluates several machine learning models, namely XGBoost, LightGBM, Gradient Boosting, and KNN, using a dataset that combines operational, temporal, and contextual features to predict food delivery times. The comparative analysis shows that tree-based approaches outperform the other models, with XGBoost achieving the best predictive performance. The findings demonstrate the strong potential of artificial intelligence and machine learning techniques for enhancing delivery efficiency and supporting sustainable urban logistics decision-making. Moreover, the proposed predictive framework can be integrated into optimization models to enhance delivery planning and support the selection of environmentally sustainable delivery modes. 
\end{abstract}
\vspace{2mm}



\section{Introduction}
The rapid expansion of food delivery services has significantly transformed urban logistics and consumer habits. Today, ordering meals via mobile applications such as \textit{Uber Eats}, \textit{Deliveroo}, or \textit{Talabat} has become part of everyday life. As consumers lead increasingly busy lifestyles, they expect fast, reliable, and on-time delivery services \cite{ahuja2021ordering}. At the same time, the growing carbon emissions generated by urban food delivery operations have intensified environmental concerns. This has led governments and logistics stakeholders to promote low-carbon delivery solutions in order to meet sustainability targets and climate mitigation objectives \cite{li2020review}. While online food delivery services offer significant convenience and economic opportunities, they also generate substantial environmental impacts, particularly through increased greenhouse gas emissions, traffic congestion, and energy consumption \cite{gray2021decarbonising}. As cities strive to achieve sustainability and carbon neutrality goals, particularly in the European Union, Canada and emerging Gulf economies such as Saudi Arabia, the transition toward low-carbon food delivery systems has become a critical challenge for researchers and policymakers \cite{bao2025future}.
Beyond environmental considerations, operational performance remains a critical concern for logistics operators and platform providers. Delivery time, in particular, represents a key performance indicator directly affecting customer satisfaction and service competitiveness \cite{li2020review}.\\
Recent advances in data analytics and machine learning (ML) offer new opportunities to enhance delivery operations. Predictive models can estimate delivery times more accurately by considering multiple factors such as distance, traffic conditions, weather, and delivery modes \cite{gurusamy2023prediction}. \\
\begin{figure*}[!t]
    \centering
    \includegraphics[width=\textwidth]{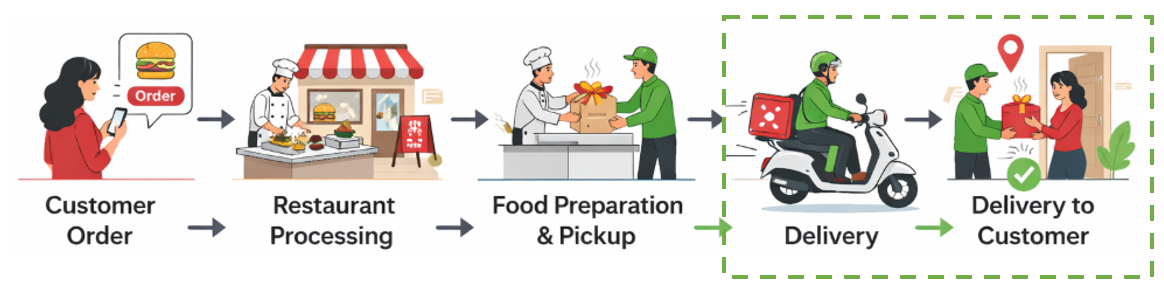}
    \caption{Online Food Delivery Process}
    \label{fig:delivery}
\end{figure*}
The objective of this study is therefore to predict food delivery times in order to evaluate and optimize low-carbon delivery strategies and support more sustainable and efficient decision-making.\\
Accordingly, this paper proposes a comprehensive study that combines a comparative analysis of low-carbon food delivery modes with a machine learning-based delivery time prediction approach: (i) reviewing and classifying existing low-carbon delivery modes; and (ii) developing and evaluating several ML models, including XGBoost, Random Forest, LightGBM, and Gradient Boosting, using real-world delivery datasets.\\

The remainder of this paper is organized as follows. Section~\ref{sec:section2} reviews the literature on low-carbon food delivery modes. Section~\ref{sec:section3} presents the dataset and preprocessing steps. Section~\ref{sec:results} describes the experimental setup, discusses the results of the ML models for delivery time prediction. Finally, Section~\ref{sec:conclusion} concludes the paper and outlines future research directions.

\section{Related Work}
\label{sec:section2}
\subsection{Online Food Delivery Systems}
\vspace{2mm}
A typical online food delivery (OFD) system involves several interconnected stages, as illustrated in Figure~\ref{fig:delivery}, the customer places an order via a mobile application, the restaurant receives and prepares the order, the courier picks it up and the meal is transported and delivered to the customer. This process highlights the complexity of coordinating multiple actors and operational factors in last-mile delivery \cite{puram2022last}.

The growth of online food delivery (OFD) platforms has been extensively studied from multiple perspectives, including optimization, consumer behavior, digitalization, and sustainability. The study in \cite{Zhen2024} developed a multi-objective optimization model to balance platform profit, rider income, and delivery efficiency by leveraging trajectory-based big data, thereby enhancing overall platform performance. Complementarily, \cite{Boldureanu2025} highlights the critical role of electronic word-of-mouth (e-WOM) as a key driver of platform adoption, using SmartPLS (Partial Least Squares Structural Equation Modeling) to demonstrate its strong influence on user behavior.

From a digitalization perspective, \cite{Gupta2024} identifies the lack of digital infrastructure as a major barrier to the adoption of Industry 4.0 technologies in food supply chains, employing Multi-Criteria Decision Making (MCDM) and barrier analysis techniques. Similarly, \cite{Annosi2023} shows that digital transformation is shifting value creation from traditional food quality toward data-driven service quality, as evidenced through a systematic literature review.

In terms of evolving customer expectations, \cite{Huq2022} proposes that order assignment strategies should account for perceived waiting time rather than solely actual travel time, using the Water Wave Optimization (WWO) algorithm to improve both customer satisfaction and operational efficiency. Furthermore, \cite{AlAdwan2023} demonstrates, through quantitative survey and regression analysis, that trust plays a mediating role between service quality and customers' intention to reuse OFD applications.


\subsection{Low-Carbon Delivery Modes}
\vspace{2mm}
The transition toward low-carbon delivery modes has become a major priority in sustainable urban logistics due to the rapid expansion of on-demand food delivery services \cite{golinska2023sustainable}. Food delivery operations are highly time-sensitive and require strict adherence to delivery deadlines in order to maintain food quality and customer satisfaction \cite{nie2025aligning}. This constraint becomes particularly critical during peak demand periods, such as lunchtime, when a large number of orders must be delivered within a very short time window \cite{ashraf2025decision}.


Several low-carbon delivery modes have recently been explored to address both environmental and operational challenges in last-mile logistics.
Electric vehicles (EVs) represent one of the most widely adopted solutions for urban deliveries due to their relatively high payload capacity and operational reliability \cite{anosike2023exploring}. However, their efficiency may decrease in dense urban environments due to traffic congestion, limited parking availability, battery constraints, and restricted curb access \cite{kim2025understanding}. In contrast, cargo bikes and electric bicycles are highly efficient for short-distance deliveries in city centers. Their ability to bypass traffic congestion and access restricted areas allows them to significantly reduce delivery times and CO$_2$ emissions \cite{betti2024uav}. Nevertheless, their limited payload capacity and rider fatigue constraints restrict their scalability during periods of high demand.

Electric scooters and motorcycles provide a balance between speed, flexibility, and energy efficiency, making them well suited for time-sensitive food delivery in dense urban areas \cite{eccarius2020powered}. For example, during lunchtime peaks, these vehicles can perform rapid point-to-point deliveries while maintaining relatively low energy consumption. However, they remain constrained by payload limitations and battery range.\\
More recently, autonomous delivery technologies such as drones and ground delivery robots have emerged as promising alternatives for low-carbon logistics \cite{belhor2025drone}. These systems can potentially reduce operational emissions and costs while improving delivery speed in specific contexts. However, their deployment is still limited by regulatory constraints, weather sensitivity, infrastructure requirements, and limited payload capacity \cite{belhor2025drone}. It can be observed that no single delivery mode can simultaneously optimize environmental impact, delivery speed, and operational reliability in urban food delivery systems. Each transportation mode offers specific advantages but also presents limitations related to capacity, accessibility, and traffic conditions. Therefore, recent studies increasingly recommend hybrid and multimodal delivery strategies that combine different modes of transport.
In such systems, EVs are commonly used to transport batches of orders from restaurants or distribution depots to urban micro-consolidation hubs, where orders are sorted and prepared for the last-mile distribution \cite{Ackva2023}. The final delivery stage is then performed by smaller and more agile vehicles such as cargo bikes or electric two-wheelers. The multimodal approach enables logistics operators to combine the high carrying capacity of EVs for medium-distance transport with the flexibility of light vehicles that can navigate dense urban areas more easily \cite{belhor2025drone}. As a result, it helps reduce congestion and CO$_2$ emissions while improving delivery efficiency and maintaining on-time delivery, particularly during peak demand periods such as lunchtime.
To further enhance operational performance and ensure service continuity in the presence of uncertainties, AI is increasingly integrated into delivery management systems \cite{moufad2025towards}. ML models can predict demand patterns, estimate delivery times, and anticipate disruptions such as traffic congestion or weather-related delays. The performance of low-carbon delivery systems is typically evaluated using several key indicators. Environmental indicators include CO$_2$ emissions and energy consumption, while operational indicators focus on delivery time, on-time delivery rate, and service reliability \textbf. Economic indicators such as operational costs and fleet use are also considered \cite{golinska2023sustainable,rodrigues2021drone}. These indicators help evaluate how efficient and sustainable last-mile food delivery systems are.

\subsection{Challenges of Sustainable Food Delivery}
\vspace{2mm}
The food delivery industry has undergone tremendous growth since the COVID-19 pandemic, raising significant sustainability challenges, particularly in achieving the United Nations 2030 Sustainable Development Goals (SDGs)~\cite{li2020review}. These challenges span multiple dimensions. From a technical perspective, battery autonomy remains a critical limitation, especially for UAVs and electric vehicles, requiring advanced scheduling algorithms, adaptive partial recharge strategies, and battery swapping mechanisms to ensure operational efficiency~\cite{betti2024uav, amiri2023bi}. In addition, the lack of robust charging infrastructure constitutes a major barrier to zero-emission delivery, motivating the development of optimization models such as mixed-integer nonlinear programming and multi-objective approaches to improve infrastructure allocation and routing with intermediate charging stations~\cite{bukhari2023zero, cokyasar2021optimization, pal2021allocation}. Vehicle driving range further constrains adoption, as highlighted by empirical studies and simulation-based or ML approaches aimed at predicting and extending EV range~\cite{gurusamy2023prediction, gandhi2026data}.

From a logistical standpoint, low-carbon delivery systems face reduced payload capacity due to battery weight, with electric trucks experiencing up to a 9\% reduction, which may be acceptable for short-range operations but remains a constraint for broader deployment~\cite{gray2021decarbonising}. Similar limitations apply to drone delivery, despite their efficiency for small parcels~\cite{bao2025future, rodrigues2021drone}. Moreover, fleet management becomes increasingly complex when dealing with geographically dispersed operations, although optimized and smart management strategies can significantly reduce emissions~\cite{zaino2024electric, lu2023emission, ferreira2025enhancing}.

Environmental and economic challenges also play a crucial role. Weather conditions and air pollution influence consumer behavior, increasing reliance on food delivery services, particularly among vulnerable populations~\cite{heldt2021cool, galati2020contribution}. At the same time, high production and distribution costs, coupled with delivery inefficiencies such as failed deliveries (estimated at 2--6\%), highlight the need for improved logistics and cost management~\cite{jurburg2023understanding, viu2020impact}. Furthermore, the transition to low-carbon delivery requires substantial upfront investments, necessitating supportive policies such as tax incentives, R\&D subsidies, and financial mechanisms like green credits~\cite{wei2024research, tian2024make}.

Finally, regulatory challenges remain a key barrier, including urban access restrictions, inadequate infrastructure, and evolving mobility policies that must be addressed to enable sustainable delivery systems~\cite{alsaleh2025electric, tiboni2021urban}. Safety and operational regulations also impact the deployment and efficiency of innovative delivery models, such as truck–drone systems, which are subject to strict regulatory frameworks despite their demonstrated potential for energy savings~\cite{eskandaripour2023last, he2024balancing}.\\

To address these challenges, it is essential to rely on data-driven approaches that capture real-world delivery conditions. Therefore,  section~\ref{sec:section3} presents the data collection and preprocessing steps, which form the basis for analyzing a delivery case study and developing ML models for delivery time prediction.

\color{black}
\section{Data Collection and Preprocessing}
\vspace{2mm}
\label{sec:section3}
The dataset used in this study was obtained from Kaggle\footnote{\url{https://www.kaggle.com/datasets/denkuznetz/food delivery-time-prediction}}. It is specifically designed for predicting food delivery times based on multiple operational and environmental factors that influence last-mile logistics performance. The dataset contains simulated yet realistic data suitable for supervised regression tasks. This dataset is highly relevant for food delivery time prediction as it combines operational variables (e.g., distance, preparation time, vehicle type, courier experience) with contextual factors such as weather, traffic level, and time of day, allowing models to reflect realistic delivery conditions. Its mix of numerical and categorical features makes it suitable for evaluating various machine learning approaches, including tree-based ensemble methods (XGBoost, LightGBM, GB) and instance-based models such as KNN. Moreover, it represents realistic urban logistics scenarios and provides a well-structured regression framework with a clearly defined target variable, making it an appropriate benchmark for performance comparison across ML algorithms.

\section{Experiments and Results}
\vspace{2mm}
\label{sec:results}
\subsection{Experimental Setup}
\vspace{2mm}
The experiments were conducted on the Food Delivery Times dataset, with the objective of predicting the delivery time in minutes. Model training was performed using Google Colab with a T4 GPU. The input features include delivery distance, traffic level, weather conditions, time of day, courier experience, and vehicle type. Data preprocessing involved removing duplicate entries and handling missing values. Categorical variables (Weather, Traffic\_Level, and Time\_of\_Day) were imputed using the mode, while the numerical feature Courier\_Experience\_yrs was imputed using the median. Categorical features were then encoded using one-hot encoding. The dataset was split into training and test sets using a 70/30 ratio with a fixed random state of 250 to ensure reproducibility.\\
Model performance was evaluated using the coefficient of determination ($R^2$) and the Mean Absolute Error (MAE). The $R^2$ score was computed on both training and test sets to assess model fit and generalization capability. MAE was used as the primary evaluation metric since it provides an interpretable measure of prediction error expressed in minutes.\\
Four regression models were evaluated: XGBoost, LightGBM, GB Regressor, and KNN. For each model, hyperparameter optimization was performed using GridSearchCV with 5-fold cross-validation on the training set. The corresponding hyperparameter search spaces are summarized in Table~\ref{tab:hyperparameter_search}.
\begin{figure*}[t]
\centering
\includegraphics[width=0.48\linewidth]{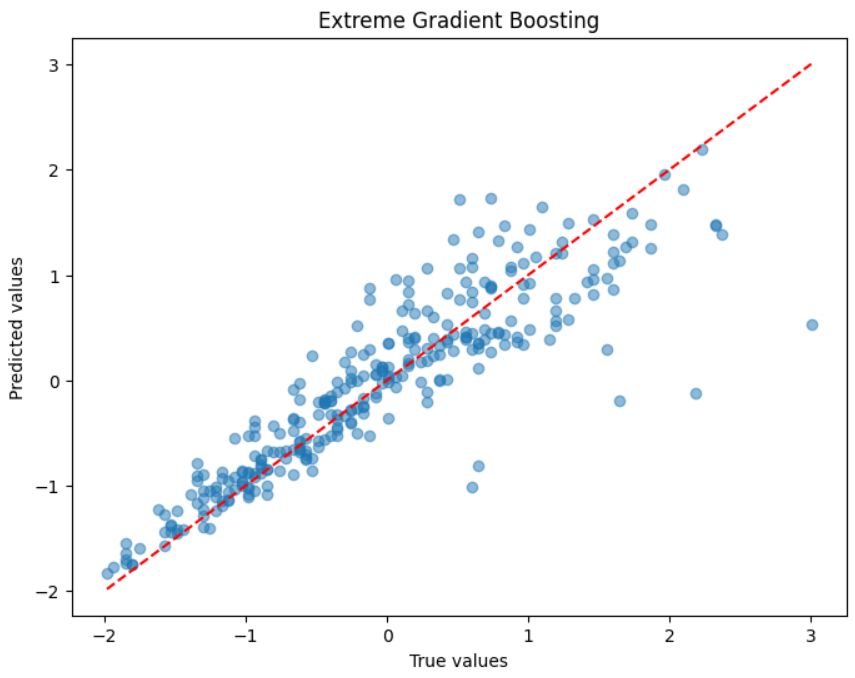}
\includegraphics[width=0.48\linewidth]{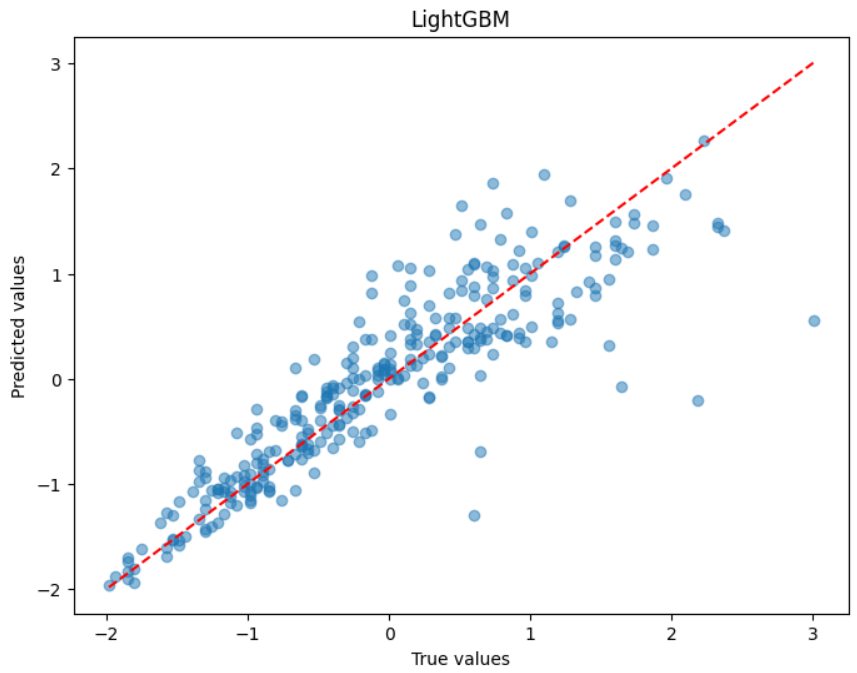}

\includegraphics[width=0.48\linewidth]{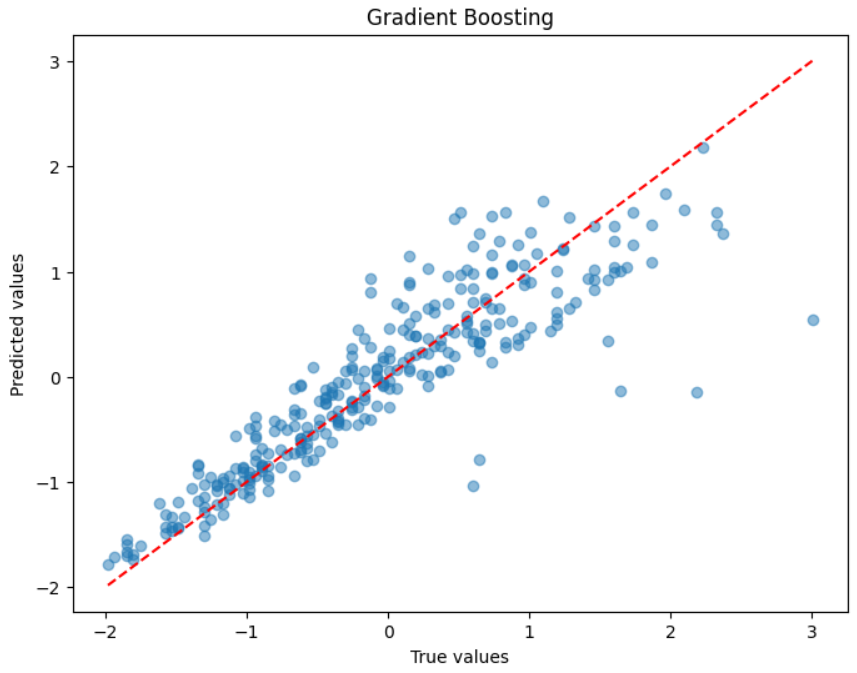}
\includegraphics[width=0.48\linewidth]{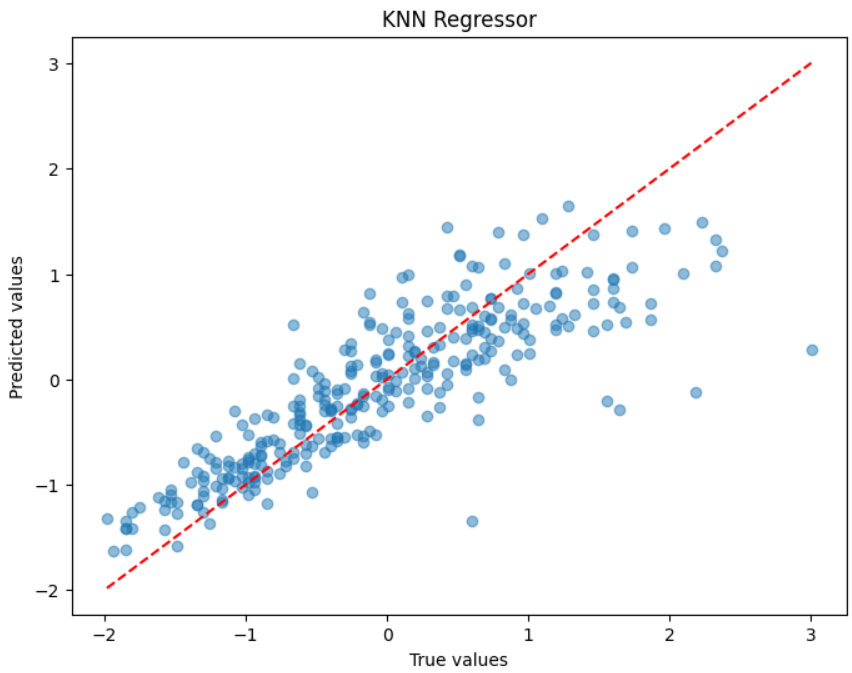}

\caption{Actual vs. predicted delivery times: (a) XGBoost, (b) LightGBM, (c) Gradient Boosting, (d) KNN.}
\label{fig:model_comparison}
\end{figure*}

\begin{table*}[h]
\caption{Hyperparameter Search Space for Each ML Model}
\centering
\begin{tabular}{lll}
\hline
\textbf{Model} & \textbf{Hyperparameter} & \textbf{Values} \\
\hline

\multirow{5}{*}{XGBoost}
& $n\_estimators$ & $\{50,100,200\}$ \\
& $learning\_rate$ & $\{0.1,0.2\}$ \\
& $max\_depth$ & $\{3,5,10\}$ \\
& $subsample$ & $\{0.8,1.0\}$ \\
& $min\_child\_weight$ & $\{1,3,5\}$ \\

\hline

\multirow{6}{*}{LightGBM}
& $n\_estimators$ & $\{50,100,200\}$ \\
& $learning\_rate$ & $\{0.05,0.1,0.2\}$ \\
& $max\_depth$ & $\{3,5,10\}$ \\
& $num\_leaves$ & $\{31,50,100\}$ \\
& $subsample$ & $\{0.8,1.0\}$ \\
& $min\_child\_weight$ & $\{1,3,5\}$ \\

\hline

\multirow{5}{*}{GB}
& $n\_estimators$ & $\{50,100,200\}$ \\
& $learning\_rate$ & $\{0.05,0.1,0.2\}$ \\
& $max\_depth$ & $\{3,5,10\}$ \\
& $subsample$ & $\{0.8,1.0\}$ \\
& $min\_samples\_split$ & $\{2,5,10\}$ \\

\hline

\multirow{4}{*}{KNN}
& $n\_neighbors$ & $\{3,5,7,10\}$ \\
& $weights$ & $\{\text{uniform},\text{distance}\}$ \\
& $p$ & $\{1,2\}$ \\
& $algorithm$ & $\{\text{auto},\text{ball\_tree},\text{kd\_tree},\text{brute}\}$ \\

\hline
\end{tabular}
\label{tab:hyperparameter_search}
\end{table*}

\subsection{Results \& Discussion}
\vspace{2mm}
\begin{table*}[t]
\centering
\caption{Model Performance Comparison}
\label{tab:model_comparison}

\renewcommand{\arraystretch}{1.4} 
\setlength{\tabcolsep}{18pt}      

\small
\begin{tabular}{lccc}
\toprule
Model & Train $R^2$ & Test $R^2$ & MAE \\
\midrule
XGBoost  & 0.8091 & 0.8019 & 0.2898 \\
LightGBM & 0.8268 & 0.7909 & 0.3003 \\
GB       & 0.8162 & 0.7979 & 0.2952 \\
KNN      & 1.0000 & 0.7909 & 0.3500 \\
\bottomrule
\end{tabular}
\end{table*}
This section presents through Table~\ref{tab:model_comparison} the performance evaluation of the proposed ML models for delivery time prediction. The models are compared using $R^2$ and MAE, and their predictive behavior is further analyzed through Figure~\ref{fig:model_comparison} of predicted versus actual values. Overall, tree-based ensemble methods outperform the instance-based KNN model in terms of accuracy and generalization. XGBoost achieves the best overall performance, with a test $R^2$ of 0.8019 and the lowest MAE (0.2898), indicating high predictive accuracy and robustness. Gradient Boosting follows closely, with a test $R^2$ of 0.7979 and MAE of 0.2952, confirming the effectiveness of boosting approaches for structured delivery data. 
In contrast, KNN exhibits clear overfitting behavior, achieving a perfect training score ($R^2 = 1.0$) but lower generalization performance, with a test $R^2$ of 0.7909 and the highest MAE (0.35). This indicates that KNN struggles to capture the underlying patterns of delivery data compared to ensemble methods.\\
Fig.~\ref{fig:model_comparison} illustrates the relationship between predicted and actual delivery times for all models. 

XGBoost shows a strong concentration of points around the diagonal, confirming its high accuracy and stability. LightGBM and Gradient Boosting also demonstrate good alignment, although with slightly higher dispersion, especially for larger delivery times. In contrast, KNN exhibits a wider spread of points and more deviation from the diagonal line, particularly for extreme values. This confirms its weaker generalization and sensitivity to data variability.\\

The experimental results clearly indicate that XGBoost provides the best trade-off between accuracy and generalization, which make it the most suitable model for real-world application. Furthermore, the superior performance of tree-based models highlights their ability to capture nonlinear relationships between delivery features such as distance, traffic conditions, weather, and courier experience. This is particularly important in the context of low-carbon delivery, where operational decisions must balance efficiency and environmental impact.






\section{Conclusion}
\label{sec:conclusion}
\vspace{2mm}
This paper proposes a ML-based approach for predicting delivery time in low-carbon food delivery systems, supported by a comprehensive review of the state of the art on online food delivery, low-carbon transport modes and associated challenges. The results show that tree-based models, particularly XGBoost, achieve high prediction accuracy, enabling improved operational efficiency and supporting more sustainable decision-making in last-mile delivery.\\

As future work, real-world case studies will be investigated to further validate the proposed framework, particularly in emerging contexts such as Saudi Arabia, where rapid urbanization and the growing demand for online food delivery create new challenges and opportunities for sustainable logistics. In addition, future research will focus on integrating the proposed ML-based prediction framework into optimization models, such as the Electric Vehicle Routing Problem (EVRP), in order to jointly optimize delivery operations, energy efficiency, and the selection of low-carbon delivery modes.

\bibliographystyle{IEEEtran}
\bibliography{References}

@inproceedings{belhor2025drone,
  title={Drone-Based Delivery in Logistics: Interdisciplinary Challenges},
  author={Belhor, Mariem and Nya, Danielle Nyakam},
  booktitle={2025 11th International Conference on Control, Decision and Information Technologies (CoDIT)},
  volume={1},
  pages={823--828},
  year={2025},
  organization={IEEE}
}

@article{li2020review,
  title={Review of online food delivery platforms and their impacts on sustainability},
  author={Li, Charlene and Mirosa, Miranda and Bremer, Phil},
  journal={Sustainability},
  volume={12},
  number={14},
  pages={5528},
  year={2020},
  publisher={MDPI}
}

@article{gurusamy2023prediction,
  title={Prediction of electric vehicle driving range and performance characteristics: A review on analytical modeling strategies with its influential factors and improvisation techniques},
  author={Gurusamy, Azhaganathan and Ashok, Bragadeshwaran and Mason, Byron},
  journal={IEEe Access},
  volume={11},
  pages={131521--131548},
  year={2023},
  publisher={IEEE}
}

@inproceedings{he2024balancing,
  title={Balancing Trade-Offs: The Energy Efficiency of Unmanned Aircraft Systems Integration in Last-Mile Delivery and Operational Policy Restrictions},
  author={He, Carrie and Singh, Parth and P’yavka, Sofiya and Wolfe, Leah and Caldwell, Kevin and Islam, Tasfia Mehbuba and Tang, Chris and Mustafa, Aws},
  booktitle={AIAA AVIATION FORUM AND ASCEND 2024},
  pages={3646},
  year={2024}
}

@article{eskandaripour2023last,
  title={Last-mile drone delivery: Past, present, and future},
  author={Eskandaripour, Hossein and Boldsaikhan, Enkhsaikhan},
  journal={Drones},
  volume={7},
  number={2},
  pages={77},
  year={2023},
  publisher={MDPI}
}

@article{tiboni2021urban,
  title={Urban policies and planning approaches for a safer and climate friendlier mobility in cities: Strategies, initiatives and some analysis},
  author={Tiboni, Michela and Rossetti, Silvia and Vetturi, David and Torrisi, Vincenza and Botticini, Francesco and Schaefer, Marco Domenico},
  journal={Sustainability},
  volume={13},
  number={4},
  pages={1778},
  year={2021},
  publisher={MDPI}
}

@article{alsaleh2025electric,
  title={Electric and Autonomous Vehicles in Italian Urban Logistics: Sustainable Solutions for Last-Mile Delivery},
  author={Alsaleh, Abdullah},
  journal={World Electric Vehicle Journal},
  volume={16},
  number={7},
  pages={338},
  year={2025},
  publisher={MDPI}
}

@article{tian2024make,
  title={Make every dollar count: The impact of green credit regulation on corporate green investment efficiency},
  author={Tian, Jinfang and Sun, Siyang and Cao, Wei and Bu, Di and Xue, Rui},
  journal={Energy Economics},
  volume={130},
  pages={107307},
  year={2024},
  publisher={Elsevier}
}

@article{wei2024research,
  title={Research on the Path of Low Carbon Transformation of Traditional Retail Enterprises Empowered by Digital Technology},
  author={Wei, Li and Xiaoxing, Qiu and Xinyu, Hu},
  journal={American Journal of Environmental Protection},
  volume={13},
  number={4},
  pages={84--92},
  year={2024},
  publisher={Science Publishing Group}
}

@article{jurburg2023understanding,
  title={Understanding the challenges facing decarbonization in the e-commerce logistics sector in Latin America},
  author={Jurburg, Daniel and L{\'o}pez, Agustina and Carli, Isabella and Chong, Mario and De Oliveira, Leise Kelli and Dablanc, Laetitia and Tanco, Mart{\'\i}n and De Sousa, Paulo Renato},
  journal={Sustainability},
  volume={15},
  number={22},
  pages={15718},
  year={2023},
  publisher={MDPI}
}

@article{viu2020impact,
  title={The impact of E-Commerce-related last-mile logistics on cities: A systematic literature review},
  author={Viu-Roig, Marta and Alvarez-Palau, Eduard J},
  journal={Sustainability},
  volume={12},
  number={16},
  pages={6492},
  year={2020},
  publisher={MDPI}
}

@article{galati2020contribution,
  title={Contribution to the sustainability challenges of the food-delivery sector: finding from the deliveroo Italy case study},
  author={Galati, Antonino and Crescimanno, Maria and Vrontis, Demetris and Siggia, Dario},
  journal={Sustainability},
  volume={12},
  number={17},
  pages={7045},
  year={2020},
  publisher={MDPI}
}

@article{heldt2021cool,
  title={Cool but dirty food?--Estimating the impact of grocery home delivery on transport and CO2 emissions including cooling},
  author={Heldt, Benjamin and Matteis, Tilman and von Schmidt, Antje and Heinrichs, Matthias},
  journal={Research in Transportation Economics},
  volume={87},
  pages={100763},
  year={2021},
  publisher={Elsevier}
}

@article{ferreira2025enhancing,
  title={Enhancing sustainable last-mile delivery: The impact of electric vehicles and AI optimization on urban logistics},
  author={Ferreira, Joao C and Esperan{\c{c}}a, Marco},
  journal={World Electric Vehicle Journal},
  volume={16},
  number={5},
  pages={242},
  year={2025},
  publisher={MDPI}
}

@article{lu2023emission
,
  title={Emission reductions from heavy-duty freight electrification aided by smart fleet management},
  author={Lu, Jiaqi and Shan, Rui and Kittner, Noah and Hu, Wenqi and Zhang, Nan},
  journal={Transportation Research Part D: Transport and Environment},
  volume={121},
  pages={103846},
  year={2023},
  publisher={Elsevier}
}

@article{zaino2024electric,
  title={Electric vehicle adoption: A comprehensive systematic review of technological, environmental, organizational and policy impacts},
  author={Zaino, Rami and Ahmed, Vian and Alhammadi, Ahmed Mohamed and Alghoush, Mohamad},
  journal={World electric vehicle journal},
  volume={15},
  number={8},
  pages={375},
  year={2024},
  publisher={MDPI}
}

@article{rodrigues2021drone,
  title={Drone flight data reveal energy and greenhouse gas emissions savings for small package delivery},
  author={Rodrigues, Thiago A and Patrikar, Jay and Oliveira, Natalia L and Matthews, H Scott and Scherer, Sebastian and Samaras, Constantine},
  journal={arXiv preprint arXiv:2111.11463},
  year={2021}
}

@article{bao2025future,
  title={The future of last-mile delivery: lifecycle environmental and economic impacts of drone-truck parallel systems},
  author={Bao, Danwen and Yan, Yu and Li, Yuhan and Chu, Jiajun},
  journal={Drones},
  volume={9},
  number={1},
  pages={54},
  year={2025},
  publisher={MDPI}
}

@article{gray2021decarbonising,
  title={Decarbonising ships, planes and trucks: An analysis of suitable low-carbon fuels for the maritime, aviation and haulage sectors},
  author={Gray, Nathan and McDonagh, Shane and O'Shea, Richard and Smyth, Beatrice and Murphy, Jerry D},
  journal={Advances in Applied Energy},
  volume={1},
  pages={100008},
  year={2021},
  publisher={Elsevier}
}

@article{gandhi2026data,
  title={A data-driven framework for estimating remaining driving range in cargo electric vehicles},
  author={Gandhi, Mrugank and Chaudhari, Archana Y and Mulla, Rahesha},
  journal={Energy Informatics},
  year={2026},
  publisher={Springer}
}

@article{pal2021allocation,
  title={Allocation of electric vehicle charging station considering uncertainties},
  author={Pal, Arnab and Bhattacharya, Aniruddha and Chakraborty, Ajoy Kumar},
  journal={Sustainable Energy, Grids and Networks},
  volume={25},
  pages={100422},
  year={2021},
  publisher={Elsevier}
}

@article{cokyasar2021optimization,
  title={Optimization of battery swapping infrastructure for e-commerce drone delivery},
  author={Cokyasar, Taner},
  journal={Computer Communications},
  volume={168},
  pages={146--154},
  year={2021},
  publisher={Elsevier}
}

@article{bukhari2023zero,
  title={Zero-emission delivery for logistics and transportation: Challenges, research issues, and opportunities},
  author={Bukhari, Janfizza and Somanagoudar, Abhishek G and Hou, Luyang and Herrera, Omar and M{\'e}rida, Walter},
  journal={The Palgrave handbook of global sustainability},
  pages={1729--1749},
  year={2023},
  publisher={Springer}
}

@article{amiri2023bi,
  title={A bi-objective green vehicle routing problem with a mixed fleet of conventional and electric trucks: Considering charging power and density of stations},
  author={Amiri, Afsane and Amin, Saman Hassanzadeh and Zolfagharinia, Hossein},
  journal={Expert Systems with Applications},
  volume={213},
  pages={119228},
  year={2023},
  publisher={Elsevier}
}

@article{eccarius2020powered,
  title={Powered two-wheelers for sustainable mobility: A review of consumer adoption of electric motorcycles},
  author={Eccarius, Timo and Lu, Chung-Cheng},
  journal={International journal of sustainable transportation},
  volume={14},
  number={3},
  pages={215--231},
  year={2020},
  publisher={Taylor \& Francis}
}

@article{betti2024uav,
  title={UAV-based delivery systems: A systematic review, current trends, and research challenges},
  author={Betti Sorbelli, Francesco},
  journal={Journal on Autonomous Transportation Systems},
  volume={1},
  number={3},
  pages={1--40},
  year={2024},
  publisher={ACM New York, NY}
}

@article{golinska2023sustainable,
  title={Sustainable urban freight for energy-efficient smart cities—systematic literature review},
  author={Golinska-Dawson, Paulina and Sethanan, Kanchana},
  journal={Energies},
  volume={16},
  number={6},
  pages={2617},
  year={2023},
  publisher={MDPI}
}

@article{nie2025aligning,
  title={Aligning sustainable development goals and stakeholders’ needs in food supply chains: a comprehensive review and mapping},
  author={Nie, Y and Hicks, B and Nassehi, A and Valero, MR},
  journal={International Journal of Production Research},
  pages={1--43},
  year={2025},
  publisher={Taylor \& Francis}
}

@article{moufad2025towards,
  title={Towards Smart and Sustainable Last Mile Delivery Systems: A Scoping Review and Conceptual Framework},
  author={Moufad, Imane and Frichi, Youness and Jawab, Fouad and Mkhalfi, Jihad},
  journal={Sustainability},
  volume={17},
  number={24},
  pages={11270},
  year={2025},
  publisher={MDPI}
  }

@article{anosike2023exploring,
  title={Exploring the challenges of electric vehicle adoption in final mile parcel delivery},
  author={Anosike, Anthony and Loomes, Helena and Udokporo, Chinonso Kenneth and Garza-Reyes, Jose Arturo},
  journal={International Journal of Logistics Research and Applications},
  volume={26},
  number={6},
  pages={683--707},
  year={2023},
  publisher={Taylor \& Francis}
}

@article{kim2025understanding,
  title={Understanding off-hour deliveries in cities: a critical review of determinants and impacts},
  author={Kim, Woojung and Calder{\'o}n, Oriana and Holgu{\'\i}n-Veras, Jos{\'e}},
  journal={Transport Reviews},
  pages={1--26},
  year={2025},
  publisher={Taylor \& Francis}
}

@article{ashraf2025decision,
  title={Decision models for order fulfillment processes of online food delivery platforms: a systematic review},
  author={Ashraf, Saad and Bardhan, Amit Kumar},
  journal={International Journal of Production Research},
  volume={63},
  number={13},
  pages={4991--5029},
  year={2025},
  publisher={Taylor \& Francis}
}

@article{Zhen2024,
  author    = {Zhen, L. and others},
  title     = {Trajectory-Based Food Delivery Order Assignment With Multi-Objective Optimization},
  journal   = {IEEE Transactions on Intelligent Transportation Systems},
  year      = {2024},
  volume    = {25},
  number    = {2},
  pages     = {1450--1463},
  publisher = {IEEE}
}

@article{Boldureanu2025,
  author    = {Boldureanu, D. and others},
  title     = {Understanding the Dynamics of e-WOM in Food Delivery Services: A SmartPLS Analysis of Consumer Acceptance},
  journal   = {Journal of Theoretical and Applied Electronic Commerce Research},
  year      = {2025},
  volume    = {20},
  number    = {1},
  pages     = {18--35},
  publisher = {MDPI}
}

@article{Gupta2024,
  author    = {Gupta, A. and others},
  title     = {Adoption of Industry 4.0 Technologies in Food Supply Chains: Barriers and Drivers},
  journal   = {Sustainability},
  year      = {2024},
  volume    = {16},
  number    = {4},
  pages     = {1542},
  publisher = {MDPI}
}

@article{Annosi2023,
  author    = {Annosi, M. C. and others},
  title     = {Digital Transformation in the Food Supply Chain: A Systematic Literature Review and Research Agenda},
  journal   = {Journal of Food Science},
  year      = {2023},
  volume    = {88},
  number    = {1},
  pages     = {12--28},
  publisher = {Wiley}
}

@article{Huq2022,
  author    = {Huq, F. and others},
  title     = {Profit and Satisfaction Aware Order Assignment for Online Food Delivery Systems Exploiting Water Wave Optimization},
  journal   = {IEEE Access},
  year      = {2022},
  volume    = {10},
  pages     = {71194--71208},
  publisher = {IEEE}
}

@article{AlAdwan2023,
  author    = {Al-Adwan, A. S. and others},
  title     = {Predicting Customer Intention to Use Online Food Delivery Services: The Role of Service Quality and Trust},
  journal   = {Administrative Sciences},
  year      = {2023},
  volume    = {13},
  number    = {3},
  pages     = {67},
  publisher = {MDPI}
}

@article{Ackva2023,
  author    = {Ackva, J. and Ulmer, M. W. and Mattfeld, D. C.},
  title     = {Consistent routing for urban micro-hubs},
  journal   = {OR Spectrum},
  year      = {2023},
  volume    = {45},
  number    = {4},
  pages     = {1117--1151},
  publisher = {Springer},
  doi       = {10.1007/s00291-023-00735-x}
}

@article{ahuja2021ordering,
  title={Ordering in: The rapid evolution of food delivery},
  author={Ahuja, Kabir and Chandra, Vishwa and Lord, Victoria and Peens, Curtis},
  journal={McKinsey \& Company},
  volume={22},
  pages={1--13},
  year={2021}
}

@article{puram2022last,
  title={Last-mile challenges in on-demand food delivery during COVID-19: understanding the riders' perspective using a grounded theory approach},
  author={Puram, Praveen and Gurumurthy, Anand and Narmetta, Mukesh and Mor, Rahul S},
  journal={The International Journal of Logistics Management},
  volume={33},
  number={3},
  pages={901--925},
  year={2022},
  publisher={Emerald Publishing Limited}
}

\end{document}